\documentclass[twocolumn,showpacs,aps,preprintnumbers,amsmath,amssymb,floatfix,prl]{revtex4}

\usepackage{color}
\usepackage{graphicx}
\usepackage{dcolumn}
\usepackage{bm}
\usepackage{multirow}
\definecolor{sienna}{cmyk}{0,0.72,1,0.45}
\definecolor{fg}{cmyk}{0.91,0,0.88,.12}
\definecolor{yellow}{cmyk}{0,0,1,0}
\definecolor{or}{cmyk}{0,1,0.5,0}
\definecolor{magenta}{cmyk}{0,1,0,0}

\definecolor{rubinered}{cmyk}{0,1,0.13,0.45}
\definecolor{blue}{cmyk}{1,1,0,0}
\definecolor{turquoise}{cmyk}{1,1,0,0.5}
\definecolor{aquamarine}{cmyk}{0,1,0,0.0}
\definecolor{midnightblue}{cmyk}{1,0.5,0.0,0.0}
\definecolor{junglegreen}{cmyk}{1,0,0.2,0.5}

\begin{document}
\title{Statistical Characterizers of Transport in Communication Networks }

\author{Satyam Mukherjee}
\affiliation{Department of Physics, Indian Institute of Technology Madras, India.}
\author{Gautam Mukherjee }
\affiliation{Bidhan Chandra College, Asansol 713304, Dt. Burdwan, West Bengal, India.}
\author{Neelima Gupte}
\affiliation{Department of Physics, Indian Institute of Technology Madras, India.}

\begin{abstract}

We identify the statistical characterizers of congestion and decongestion for 
message transport in model communication lattices. These turn out to be the travel time distributions, which are Gaussian in the congested phase, and log-normal in the decongested phase. Our results are demonstrated for two dimensional lattices, such the Waxman graph, and for lattices with local clustering and geographic separations, gradient connections, as well as for a $1-d$ ring lattice with random assortative connections. 
The behavior of the distribution identifies the congested and decongested phase correctly for these distinct network topologies and decongestion strategies. 
The waiting time distributions of the systems also show identical signatures of 
the congested and decongested phases.
\end{abstract}

\maketitle

Investigations of traffic flows on substrates of various topologies have been a topic of recent research interest. \cite{tadic}.
Congestion  effects can occur in real networks like telephone networks, computer networks and the Internet  due to various factors like capacity, band-width and network topology \cite{Huang}. These lead to deterioration of the service quality experienced  by users  due to an increase in network load.
Statistical characterizers which can identify the state of the network, whether congested or decongested, can be of practical utility. In this paper, we identify statistical characterizers which carry the signature
 of the state of congestion or decongestion of the network.

The statistical characterizer which carries the signature of the congested or decongested phase, is identified to be the travel time distribution of the messages. The travel time distribution has been studied earlier in the context of vehicular traffic \cite{Nagatani}, server traffic \cite{Olsen} and the Internet \cite{Kang}.  Hence the travel time distribution can be regarded as an useful statistical characterizer of transport. In our model networks, the travel time is defined to be the time required for a message to travel from source to target, including the time spent waiting at congested hubs.
This distribution turns out to be normal or Gaussian in the congested phase, and log-normal in the decongested phase.  

We demonstrate that the travel time distribution is able to identify correctly the congested/decongested state in the case of two  dimensional model networks, such as the Waxman topology network, a popular model for  Internet topology\cite{waxmangraph}, as well as for a network with local clustering\cite{braj1}, and its variants with gradient connections \cite{sat}. The same characterizer is able to distinguish between the congested and decongested phases in a network with a one dimensional ring geometry. Thus, the travel time distribution is a robust characterizer of the congested/decongested phase.

We first consider models based on $2-d$ lattices. We note that communication networks based on two-dimensional lattices have been considered earlier in the context of search algorithms \cite{kleinberg} and of network traffic with routers and hosts \cite{Ohira,Sole} and have been observed to reproduce realistic features of Internet traffic. 

\begin{figure*}
\begin{center}
\begin{tabular}{cccccc}
(a)&
\includegraphics[height=5cm,width=5.25cm]{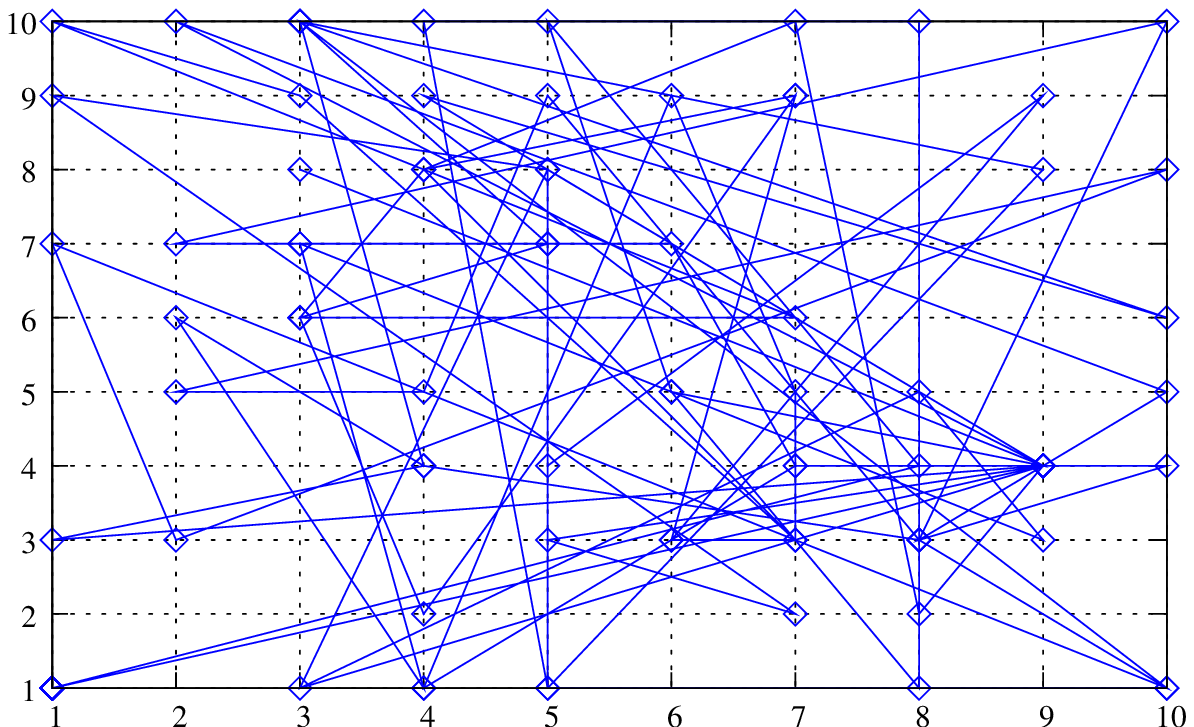}&
(b)&
\includegraphics[height=5cm,width=5.25cm]{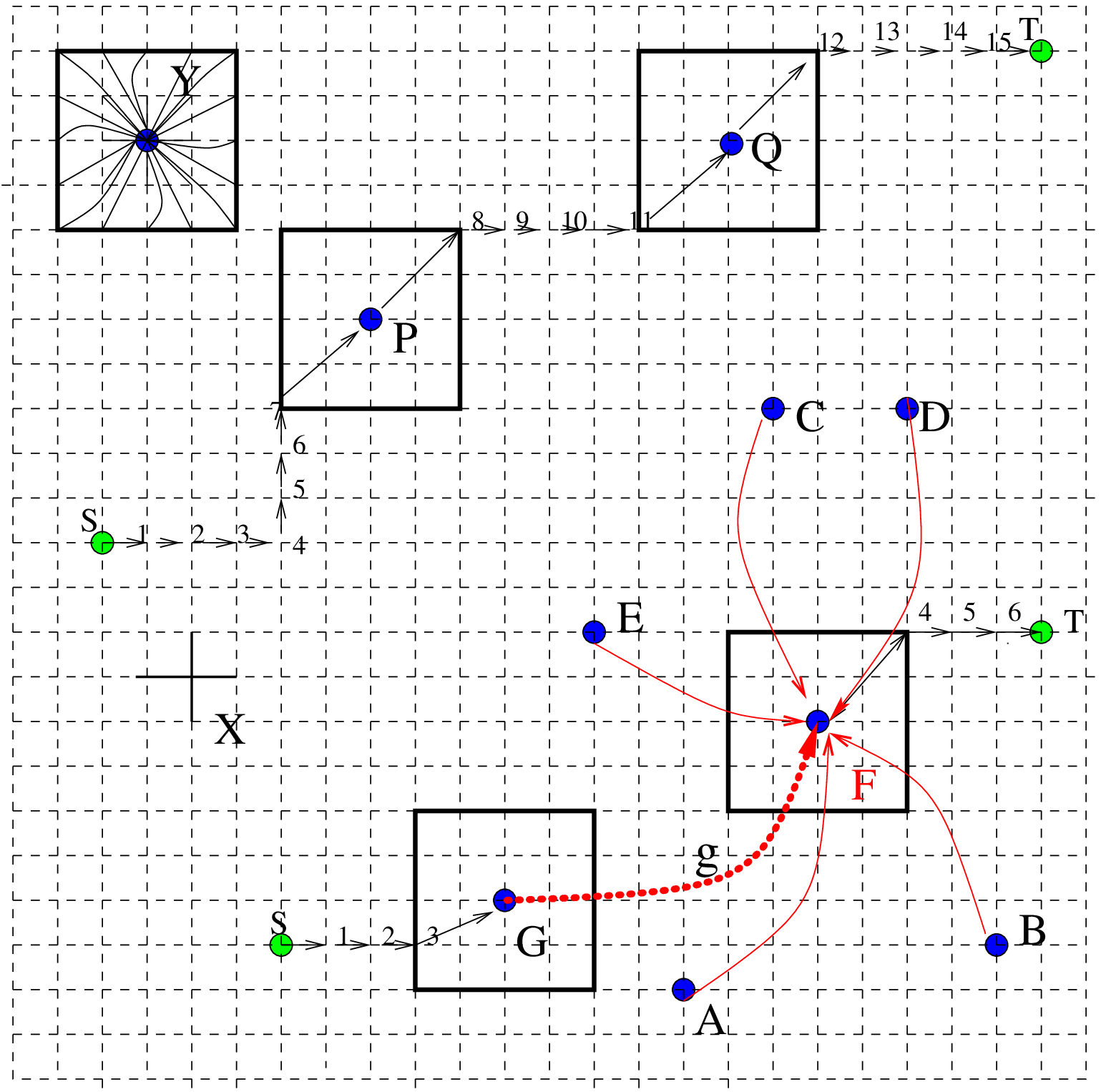}&
(c)&
\includegraphics[height=5cm,width=5.25cm]{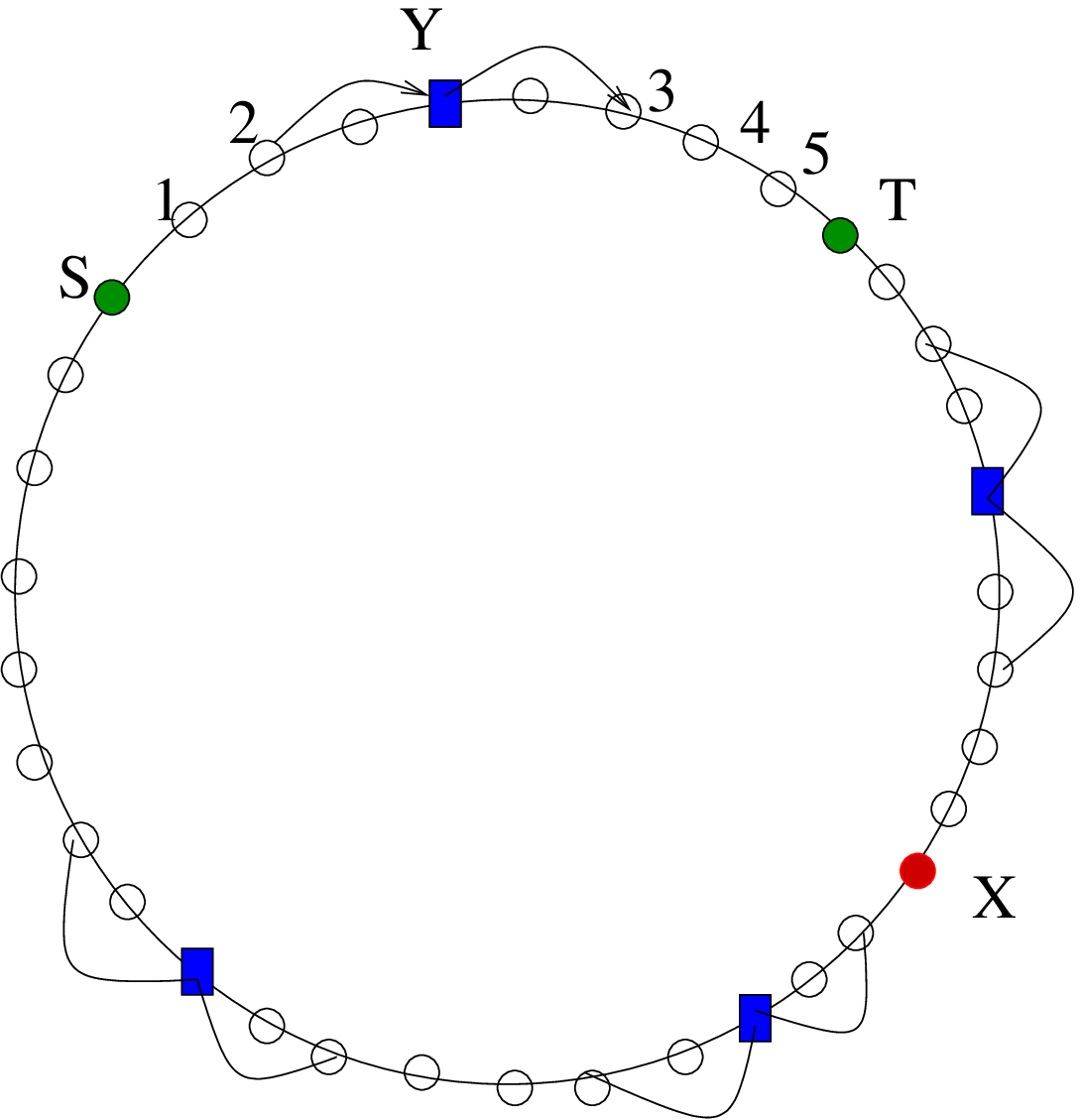}\\
\end{tabular}
\end{center}

\caption{\label{fig:qlairp}(a) The figure shows a Waxman topology network generated by connecting $55$ points by the Waxman algorithm for $\alpha=0.05$ and $\beta=0.1$, on a $10\times 10$ lattice.  The number of links increases as the values of $\alpha$ and $\beta$ are increased. (b) A regular two dimensional lattice. {\it X} is an ordinary node with nearest neighbor connections. Each hub has a square influence region (as shown for the hub {\it Y}). A typical path from the source $S$ to the target $T$ is given by the path S-1-2-3-$\cdots$-7-P-8-$\cdots$-11-Q-12-$\cdots$-T. After the implementation of the gradient mechanism, the distance between $G$ and $F$ is covered in one step as shown by the link $g$ and a  message is routed along the path $S-1-2-3-$G$-$g$-$F$-4-5-6-T$. (c) A $1-d$ ring lattice of ordinary nodes ($X$) with nearest neighbor connections and randomly distributed hubs ($Y$). 
 h rough the hub $Y$. 
}
\end{figure*}

The first network based on a $2-d$ geometry is the Waxman graph \cite{waxmangraph}, which incorporates  the distance dependence in link formation which is characteristic of real world networks \cite{lakhina} and  has  been widely used to model the topology of intra-domain networks \cite{verma}.  We consider the case where the Waxman graphs are generated on a rectangular coordinate grid of side $L$ with the probability $P(a,b)$ of an edge from node $a$ to node $b$ given by
\begin{equation}
P(a,b)=\beta\exp(-\frac{d}{\alpha M})
\end{equation}
where the parameters $0< \alpha,\beta < 1$, $d$ is the Euclidean distance from $a$ to $b$ and $M=\sqrt2\times L$ is the maximum distance between any two nodes \cite{waxmangraph}. Large values of $\beta$ result in graphs with larger link densities and small values of $\alpha$ increase the density of short links as compared to the longer ones. A topology similar to Waxman graphs  is generated by selecting randomly a predetermined number $N_{w}$ of nodes in the $2-d$ lattice for generating the edges. Additionally, each node of the lattice has a connection to its nearest neighbors ( See Fig.\ref{fig:qlairp}(a)). 

The second network that we study is a model which incorporates local clustering and geographic separations developed in Ref.\cite{braj1}. As shown in Fig.\ref{fig:qlairp}(b), this network consists of  a $2-d$ lattice with nodes and hubs, where the hubs are randomly located on the lattice, and are  connected to all nodes inside their given area of influence \cite{footn1}.

\begin{figure}[!t]
\begin{center}
\includegraphics[height=5.25cm,width=7.25cm]{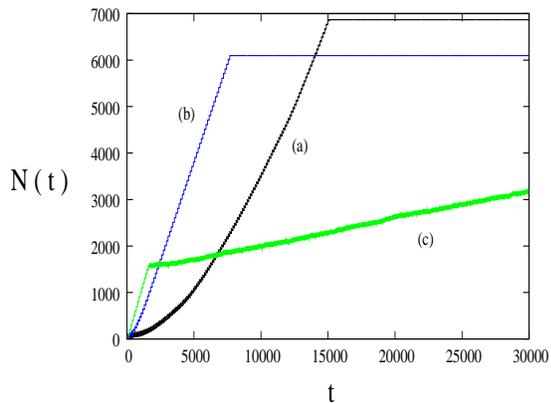}
\end{center}
\caption{\label{fig:model1d2d}(Color online) The plot of number of messages $N(t)$ flowing on the lattice as a function of time $t$ for (a) the Waxman topology network and the baseline mechanism for (b) the locally clustered $2-d$ lattice and (c) the $1-d$ ring network. \\}
\end{figure}

A given number of messages $N_m$ are allowed to travel on these lattices between fixed source target pairs by a distance based routing algorithm by which a node which holds a message looks for a hub in the direction of the target which is nearest to itself, and routes the message to it. (See Fig. 1(a) for a typical path). When many messages travel on the network, the finite capacity of the hubs can lead to the  trapping of messages in their neighborhoods, and a consequent congestion or jamming of the network. Here, we study a situation where $N_m$ messages are deposited at regular intervals on the network. If the message deposition takes place faster than the rate at which messages clear, the network can congest \cite{footn2}. We plot the number of messages $N(t)$ which are flowing on both the $2-d$ lattices as a function of time $t$ as shown in Fig.\ref{fig:model1d2d}. We allow $N_{m}=100$ messages to run continuously at every $120$ time steps for a given run time. For these values, the networks get congested and $N(t)$ gets saturated indicating formation of transport traps as seen in \cite{sat}. The reasons for trapping include the opposing movement of messages from sources and targets situated on different sides of the lattice, as well as edge effects.
\begin{figure*}
\begin{tabular}{cccccc}
(a)&
\includegraphics[height=5.25cm,width=5.25cm]{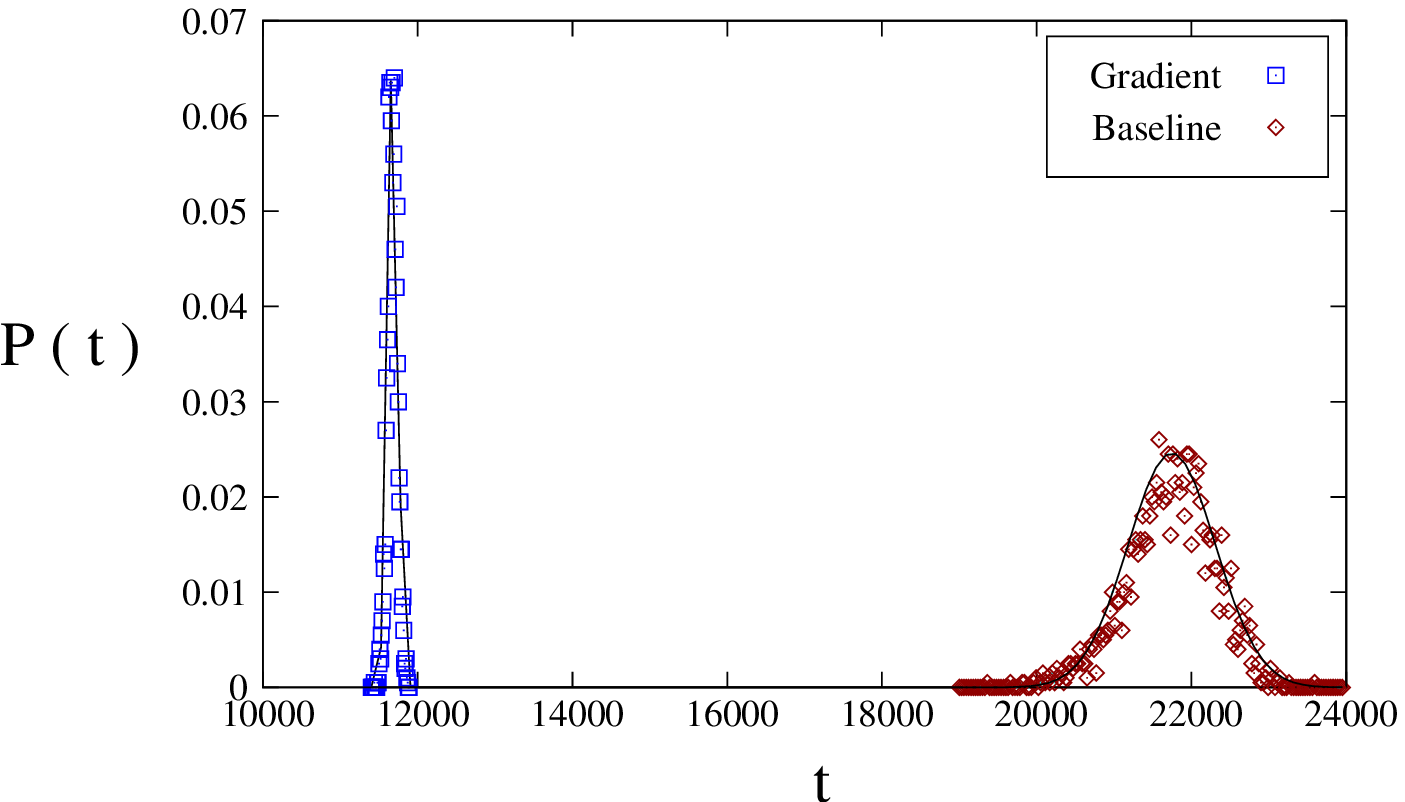}&
(b)&
\includegraphics[height=5.25cm,width=5.25cm]{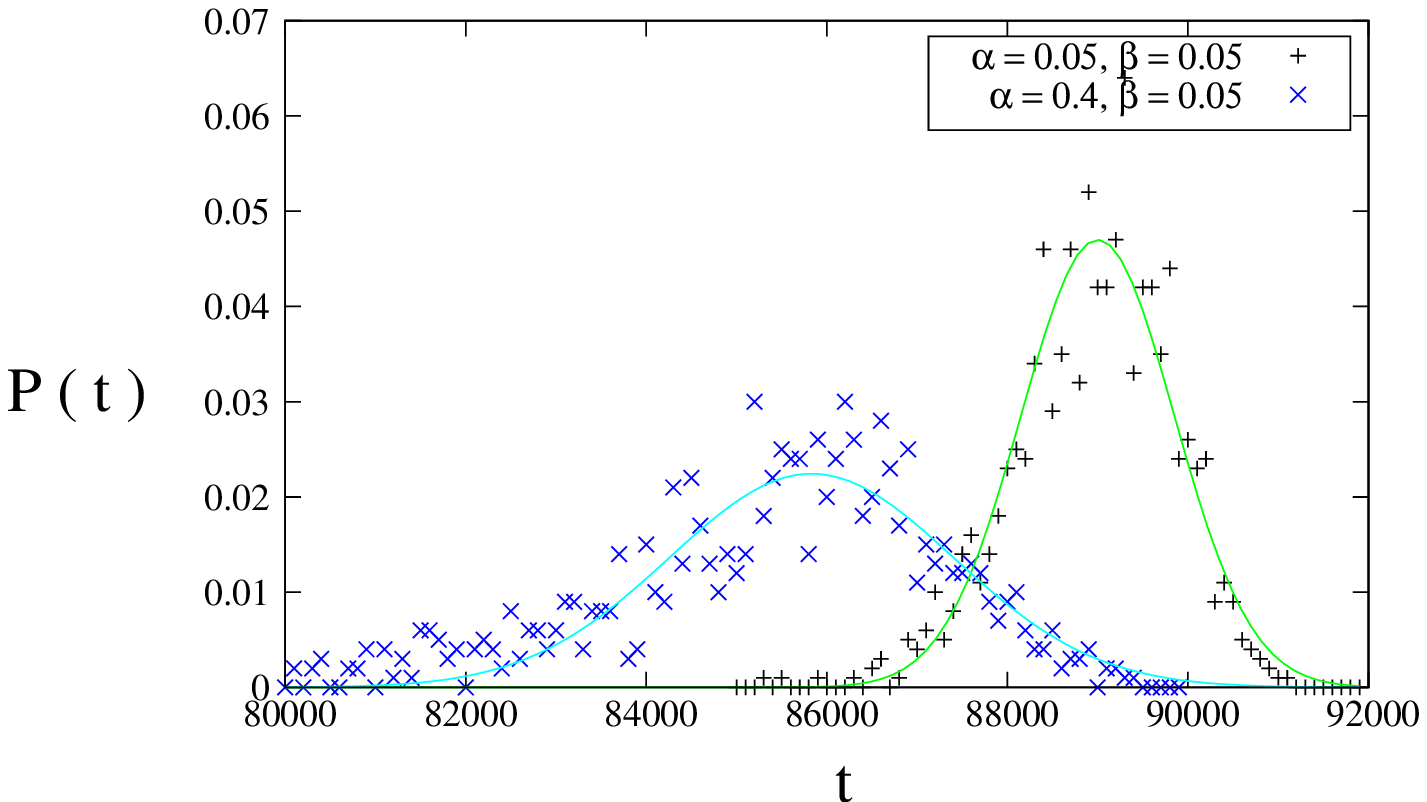}&
(c)&
\includegraphics[height=5.25cm,width=5.25cm]{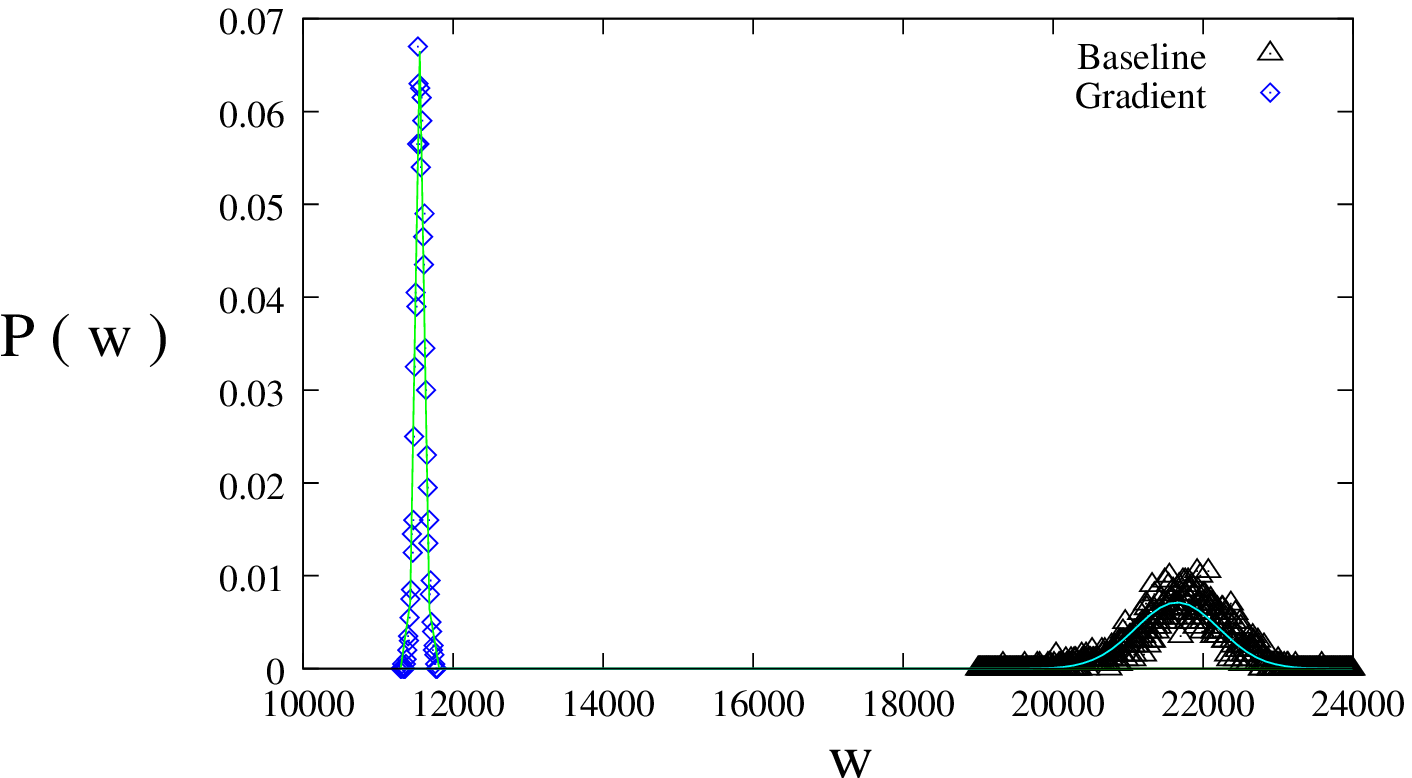}\\
(d)&
\includegraphics[height=5.25cm,width=5.25cm]{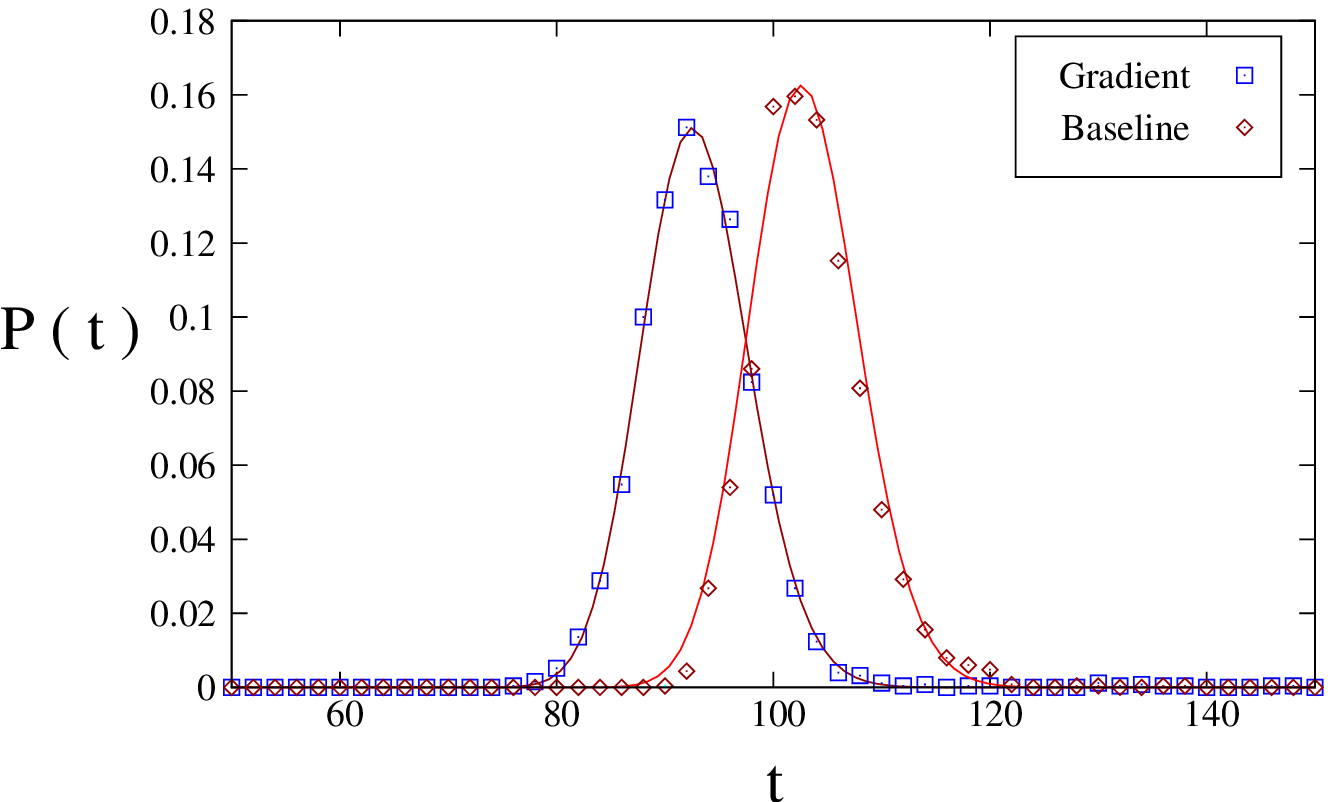}&
(e)&
\includegraphics[height=5.25cm,width=5.25cm]{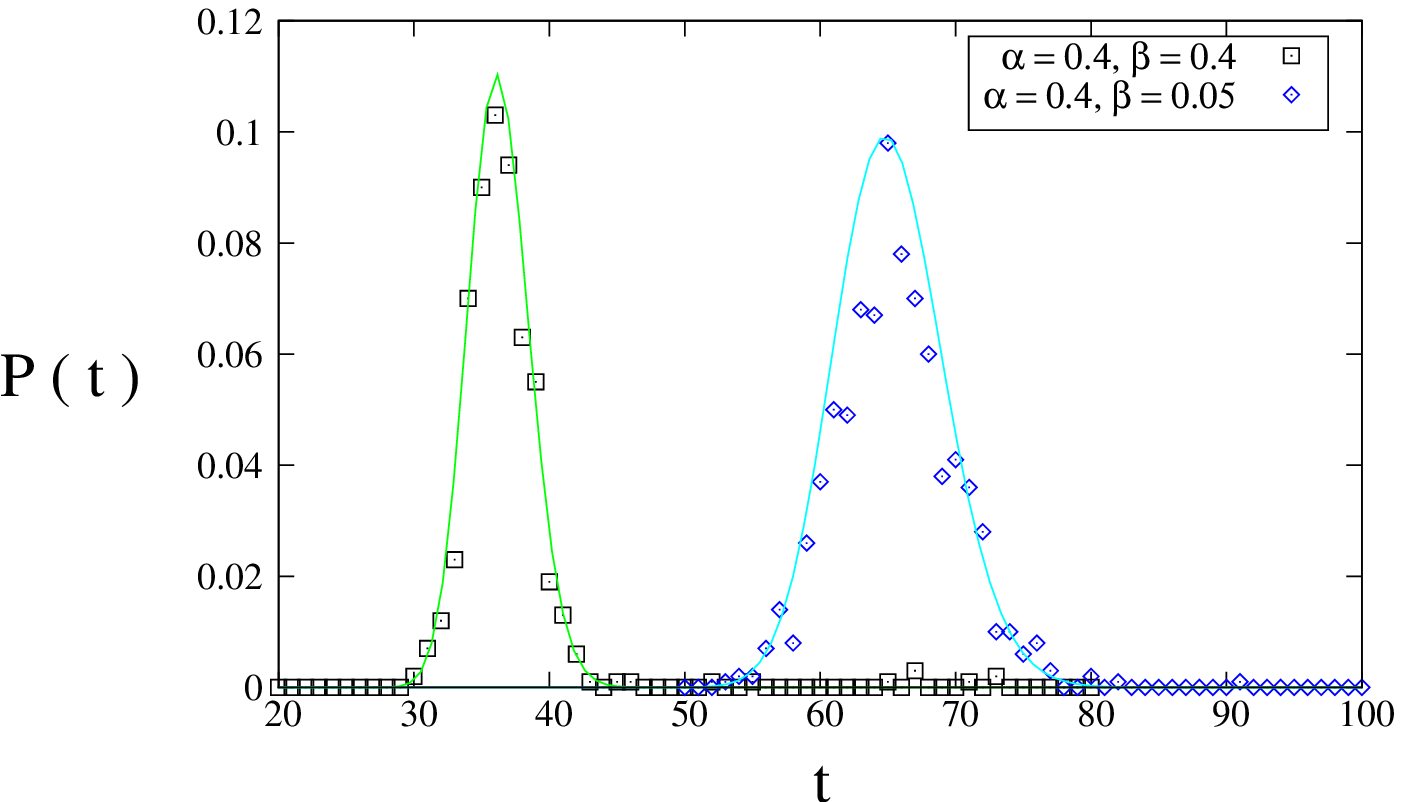}&
(f)&
\includegraphics[height=5.25cm,width=5.25cm]{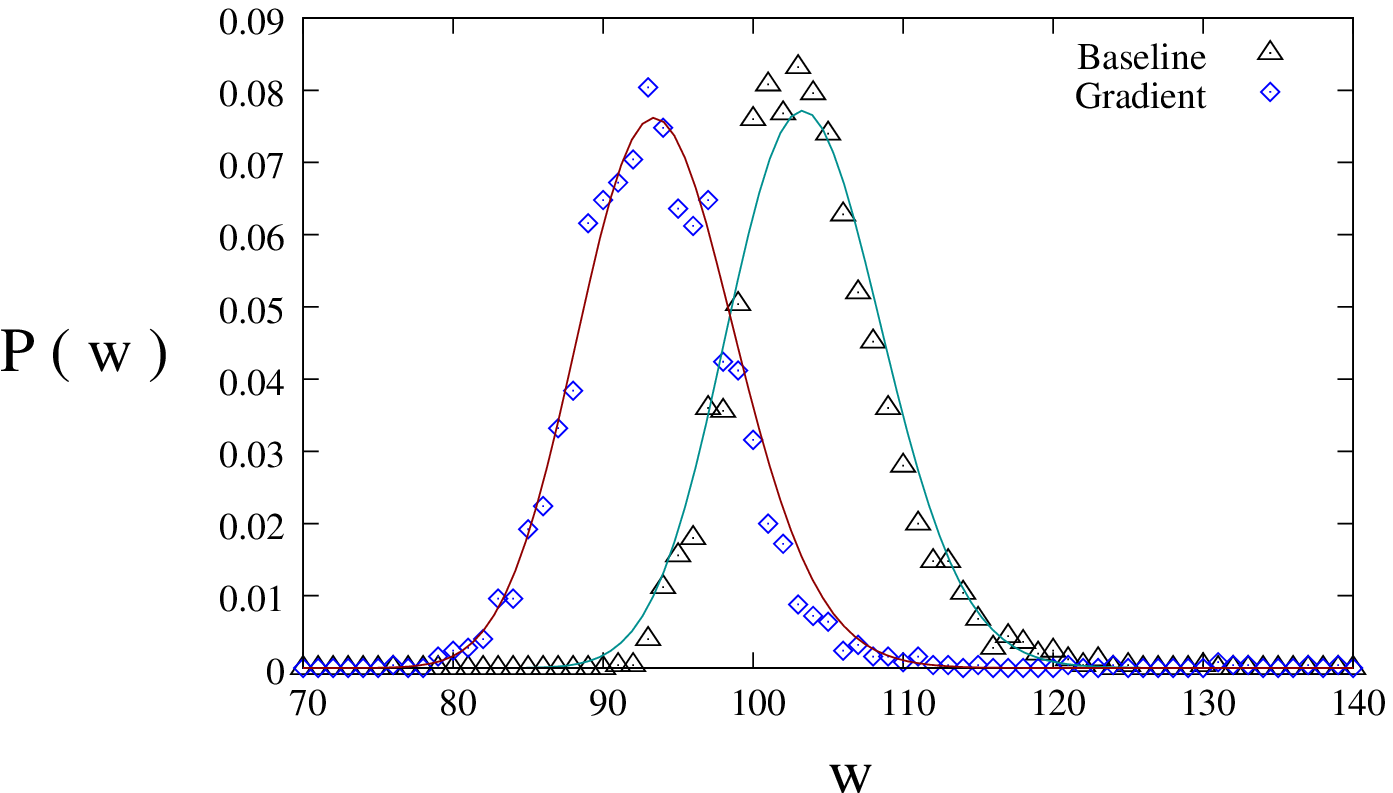}\\
\end{tabular}
\caption{\label{fig:trap2} For (a), (b), (c) the travel time and waiting time distributions in the congested phase shows a Gaussian distribution. For (d), (e), (f) the distributions change to a log-normal in the decongested phase. (a) $\sigma$ (gradient) = $64.07$ ($\chi^{2}=0.026$) and $\sigma$ (baseline) = $567.31$ ($\chi^{2}=0.195$). (b) $\sigma$ ($\alpha=0.05, \beta=0.05$) = $850.37$ ($\chi^{2}=0.463$) and $\sigma$ ($\alpha=0.4, \beta=0.05$) = $1576$ ($\chi^{2}=0.852$). (c) $\sigma$ (gradient) = $60$ ($\chi^{2}=0.02$) and $\sigma$ (baseline) = $560$ ($\chi^{2}=0.2$). (d) $\sigma$ (gradient)= $0.122$ ($\chi^{2}=0.075$) and $\sigma$ (baseline) = $0.113$ ($\chi^{2}=0.075$). (e) $\sigma$ ($\alpha=0.4, \beta=0.4$) =  and $0.062$ ($\chi^{2}=0.06$) and $\sigma$ ($\alpha=0.4, \beta=0.05$) = $0.06$ ($\chi^{2}=0. 066$). (f) $\sigma$ (gradient) = $0.05$ ($\chi^{2}=0.04$) and $\sigma$ (baseline) = $0.056$ ($\chi^{2}=0.04$). Here $\sigma$ is the standard deviation and $\chi^{2}$ is the chi-squared test for accuracy of the fit.\\}
\end{figure*}

We can now identify the statistical characteriser of the congested and decongested phase for these two networks. This turns out to be the travel time distribution. Here the travel time is the total travel time of messages including the time each message waits on all the nodes to be delivered to adjacent node along the path of their journey to respective targets.  
For the Waxman topology network if messages are fed on the system at a constant rate of $N_{m}=100$ messages at every $120$ time steps for a total run time of $90000$, messages are not delivered to their targets and the network is in the congested phase. For the locally clustered $2-d$ network we allow $N_{m}=100$ messages be deposited  at every $120$ time steps in a $100\times 100$ lattice and $D_{st}$ = 142, for $100$ hubs and total run time of $60000$.  At this value of $N_{m}$ many messages remain undelivered in the lattice due to the onset of traps and the system is in the maximal congested regime. The travel time distribution for this congested phase for both these networks is shown in Fig.\ref{fig:trap2}. The travel time distribution can be fitted by a Gaussian of the form
\begin{equation}
P(t)=\frac{1}{{\sigma}{\sqrt {2\pi}}}\exp(-\frac{(t-\mu)^{2}}{2{\sigma}^2})
\end{equation} 
If $100$ messages are fed continuously at every $200$ time steps all the messages get delivered to their targets for both cases, and the data for the travel time distribution can be fitted by a log-normal distribution of the form
\begin{equation}
P(t)=\frac{1}{t{\sigma}{\sqrt {2\pi}}}\exp(-\frac{({\ln}t-{\mu})^{2}}{2{\sigma}^{2}})
\end{equation}   
\begin{figure}
\begin{center}
\begin{tabular}{cc}
\hspace{-0.4cm}
\includegraphics[height=5.25cm,width=4.25cm]{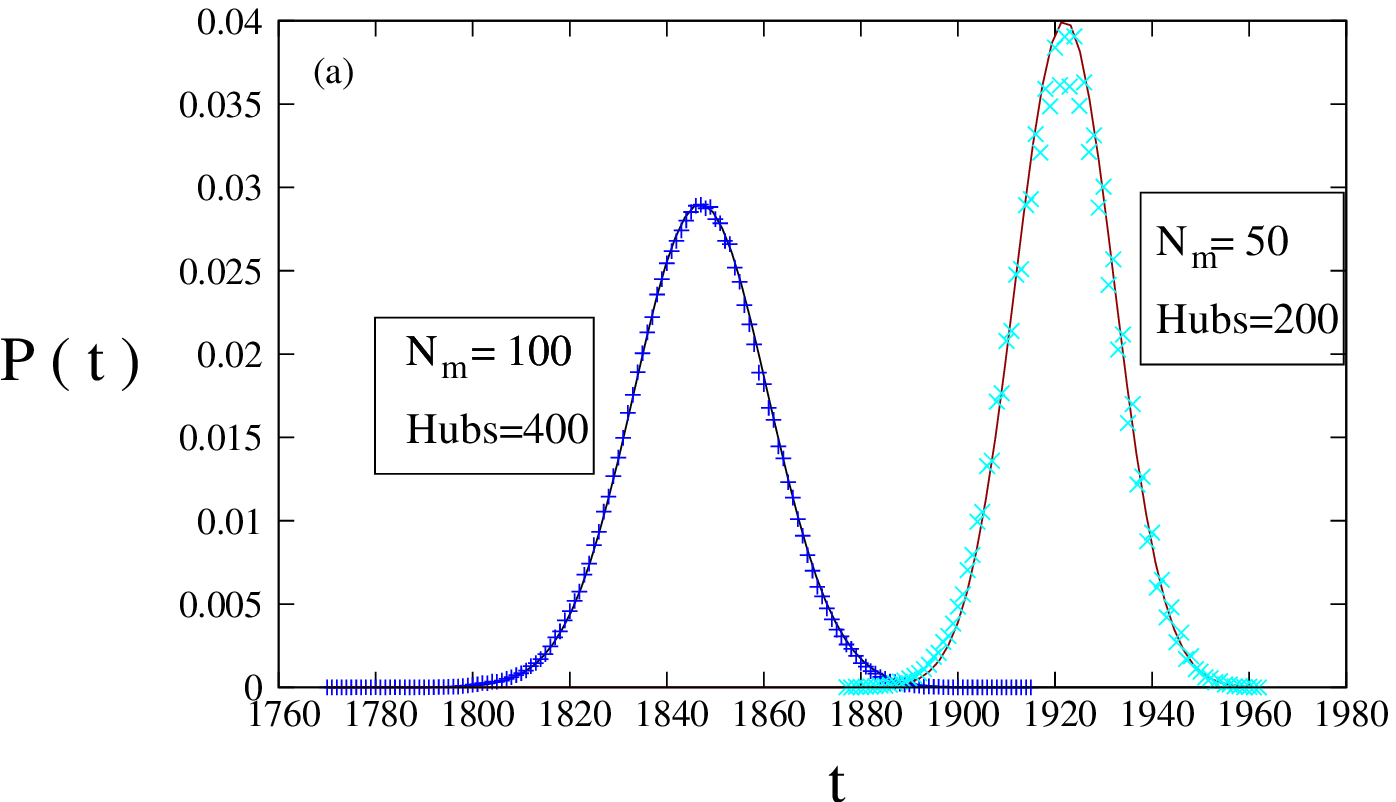}&  
\includegraphics[height=5.25cm,width=4.25cm]{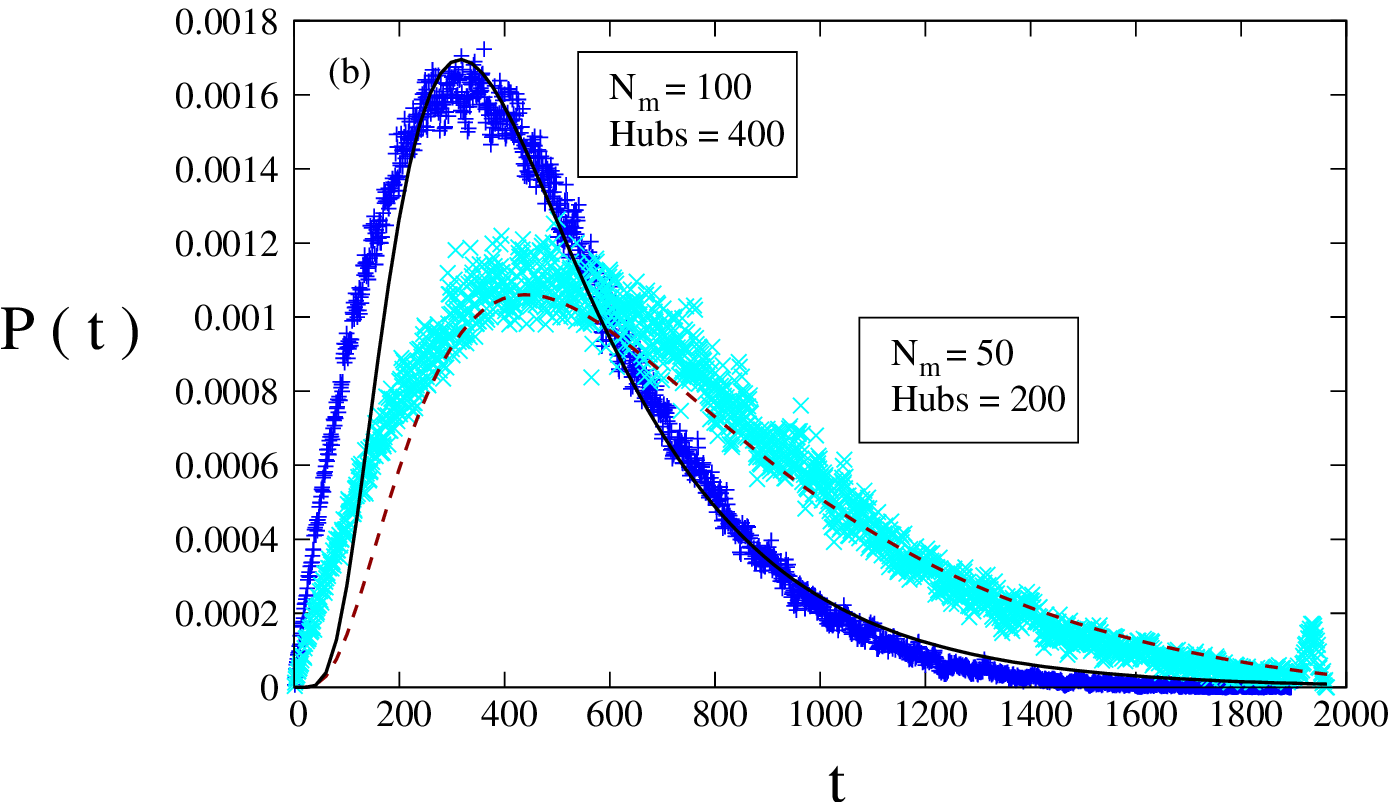}\\  
\end{tabular}
\end{center}
\caption{\label{fig:wait1}(Color online) The plot of travel time distribution of messages shows (a) Gaussian distribution in the congested phase. The standard deviation $\sigma$ is $(i)$ 13.86 ($\chi^{2}=0.0086$) for $N_{m}=100$ and 400 hubs $(ii)$ 10.21 ($\chi^{2}=0.0095$) for $N_{m}=50$ and 200 hubs. (b) Log-normal behavior with a power law correction is seen in the decongested phase. $(i)$ $N_{m}=100$ and 400 hubs, $\sigma=1.42$ ($\chi^{2}=0.14$), $\beta=0.88$, $B=-0.0009$ and $(i)$ $N_{m}=50$ and 200 hubs, $\sigma=1.79$ ($\chi^{2}=1.75$), $\delta=0.91$, $B=-0.0009$.\\}
\end{figure}

Thus it is evident that during the congested phase the travel time distribution for messages traveling at constant density in the network, shows  Gaussian behavior (Fig.\ref{fig:trap2}(a) and Fig.\ref{fig:trap2}(b)). On the other hand log-normal behavior is found during the decongested phase (Fig.\ref{fig:trap2}(d) and Fig.\ref{fig:trap2}(e)).

An efficient way of decongesting the lattice has turned out to be  the gradient mechanism \cite{sat}. This  is implemented by identifying the hubs with the five highest values of $CBC$\cite{footn3}, assigning them capacity proportional to their CBC values, and setting up a gradient to the hub with the highest capacity.
The decongestion/congestion transition occurs at a much higher value once the 
gradient strategy is implemented. Here again, the decongested phase shows a log-normal distribution of travel times, and the normal phase shows a gaussian distribution of travel times. Similar results are seen for other decongesting strategies, such as connecting the hubs of high CBC by random assortative connections
\cite{braj1}. The waiting time distribution of the system, where the waiting time is defined to be the time for which the messages wait at congested nodes, also show an identical signature of the congestion/decongestion transition (Fig.\ref{fig:trap2}(c) and Fig.\ref{fig:trap2}(f)).


We also propose an one dimensional version of the communication network of nodes and hubs. The base network is a ring lattice of size $L$ with nearest neighbor interaction. Hubs are distributed randomly in the lattice where each hub has $2k$ nearest neighbors (Fig.\ref{fig:qlairp}(c)) \cite{footn1}. 
In our simulation we have taken $k$=4, although Fig.\ref{fig:qlairp}(c) illustrates only $k=2$ connections. The distance between a source and target is defined by the Manhattan distance $D_{st}=|is -it|$. If a message is routed from a source $S$ to a target $T$ on this lattice through the baseline mechanism, it takes the path $S-1-2-Y-3-4-5-T$ as in  Fig.~\ref{fig:qlairp}(c). The routing algorithm is same as that used in the $2-d$ model \cite{braj1,sat}. A given number $N_{m}$ of source and target pairs start sending $N_{m}$ messages continuously at every $100$ time steps for a total run time of $30000$. The plot of $N(t)$ as a function of time $t$ for this lattice (Fig.\ref{fig:model1d2d}(c)) attains saturation for $t\simeq 10^{8}$.

The  travel time is calculated for a source-target separation of  $D_{st}=2000$ on a $L=10000$ ring lattice, and averaged over $1000$ hub realizations. For the baseline mechanism, where network congests at these values, the data for travel time distribution can be fitted by a Gaussian (Fig.\ref{fig:wait1}(a)). If the hubs are connected by the assortative mechanism, all the messages clear, and  the distribution can be fitted well by a log-normal function with a power law correction of the form
\begin{equation}
P(t)=\frac{1}{t{\sigma}{\sqrt {2\pi}}}\exp(-\frac{({\ln}t-{\mu})^{2}}{2{\sigma}^{2}})(1+Bt^{-\delta})
\end{equation}
as shown in Fig.\ref{fig:wait1}(b).                                       

Thus if the hubs are connected by assortative mechanisms, there is no congestion, and the leading behavior is log-normal as in the decongested case of the $2-d$ networks. An additive power law correction is seen due to the $1-d$ nature of the network. Due to the ring geometry of the network, some messages are not routed through the links created due to the assortative connections between hubs.  These messages thus have larger travel times, and contribute the additive power-law corrections to the basic log-normal behavior in the decongested phase.        

To summarize, the statistical characterizers of the communication networks studied here, viz. the travel time distributions show the characteristic signatures of the congested or decongested state of the network being normal in the congested phase and log-normal in the decongested phase. The results are true for the locally clustered  communication network as well as the Waxman topology network, and also carry over to a one-dimensional lattice, to leading order. Thus the travel time distribution is a robust characterizer of the congested or decongested phase. For the $1-d$ lattice, the distribution carries an additional signature of the topology in the form of an additive power law correction to the leading order. The waiting time distributions carry  identical signatures of the congested and decongested phase. These results are valid for different lattice sizes and hub densities as well \cite{footn4}. We note that networks which incorporate geographic clustering and encounter congestion problems arise in many practical situations e.g. cellular networks\cite{Jeong} and air traffic networks \cite{Sinai}. It would be interesting to see if our results have relevance in real life  contexts.

We thank CSIR, India for support under their extra-mural scheme.


\begin{thebibliography}{90}
\bibitem{footn1} The hubs are separated by a minimum distance $d_{min}=1$ and nodes in the overlap areas of the hubs acquire connections to all the hubs which overlap.
\bibitem{footn2}Networks can congest even with a one time deposition of messages, if the number of messages is sufficiently high. All the subsequent discussion here is applicable for this case as well.
\bibitem{footn3}The CBC of the $m^{th}$ hub is defined as the ratio $CBC={N_m}/N$ where $N_m$ is the number of messages that pass through the hub $m$, and $N$ is the total number of messages running on the lattice.
\bibitem{footn4} We observed similar behavior for $(i)$ $L=5000$, $N_{m}=50$, $D_{st}=1000$ and $200$ hubs and $(ii)$ $L=20000$, $N_{m}=200$, $D _{st}=4000$ and $800$ hubs for the $1-d$ lattice.
\end{thebibliography}
\end{document}